\documentstyle[seceq,preprint,eclepsf]{jpsj}

\newcommand{\PR}{{\rm Phys.\ Rev.}}
\newcommand{\PRL}{{\rm Phys.\ Rev.\ Lett.}}
\newcommand{\JPSJ}{{\rm J.\ Phys.\ Soc.\ Jpn.}}
\newcommand{\PTP}{{\rm Prog. Theor. Phys.}}
\newcommand{\SSC}{{\rm Solid State Commun.}}
\newcommand{\SM}{{\rm Synth. Met.}}
\newcommand{\eql}{ < \kern -12pt \lower 5pt \hbox{$\displaystyle =$}}
\newcommand{\eqg}{ > \kern -12pt \lower 5pt \hbox{$\displaystyle =$}}
\newcommand{\la}{ < \kern -12pt \lower 5pt \hbox{$\displaystyle \approx$}}
\newcommand{\ga}{ > \kern -12pt \lower 5pt \hbox{$\displaystyle \approx$}}
\newcommand{\lsim}{{ < \kern -11.2pt \lower 4.3pt \hbox{$\displaystyle \sim$}}}
\newcommand{\gsim}{{ > \kern -11.2pt \lower 4.3pt \hbox{$\displaystyle \sim$}}}
\newcommand{\del}[2]{\frac{\partial{#1}}{\partial{#2}}}
\newcommand{\bvec}[1]{\mbox{\boldmath$#1$}}

\newcommand{\vare}{\varepsilon}
\newcommand{\EF}{\varepsilon_{\mbox{\tiny F}}}

\newcommand{\VF}{v_{\mbox{\tiny F}}}
\newcommand{\tauz}{\tau_{\mbox{\tiny{0}}}}
\newcommand{\taue}{\tau_\varepsilon}
\newcommand{\vecq}{\bvec{q}}
\newcommand{\Tr}{\mbox{\rm Tr}}
\newcommand{\Dpara}{D_{\mbox{\tiny $\|$}}}
\newcommand{\Dperp}{D_{\mbox{\tiny $\bot$}}}
\newcommand{\Spara}{\sigma_{\mbox{\tiny $\|$}}}
\newcommand{\Sperp}{\sigma_{\mbox{\tiny $\bot$}}}
\newcommand{\Le}{L_{\vare}}
\newcommand{\Lepara}{L_{\vare \mbox{\tiny $\|$}}}
\newcommand{\Leperp}{L_{\vare \mbox{\tiny $\bot$}}}
\newcommand{\I}{\mbox{\rm \large I}}

\def\runtitle{Weak Field Magnetoresistance in Quasi-One-Dimensional Systems}
\def\runauthor{Yoshitaka {\sc Nakamura} and Hidetoshi {\sc Fukuyama}}

\title
{
Weak Field Magnetoresistance in Quasi-One-Dimensional Systems 
}

\author
{ 
Yoshitaka {\sc Nakamura}
and Hidetoshi {\sc Fukuyama}
}

\inst
{
Department of Physics, Faculty of Science, University of Tokyo, Tokyo 113 
}

\recdate
{
December 15, 1997
}

\abst { Theoretical studies are presented on weak localization effects
and magnetoresistance in quasi-one-dimensional systems with open Fermi
surfaces. Based on the Wigner representation, the magnetoresistance in
the region of weak field has been studied for five possible
configurations of current and field with respect to the
one-dimensional axis. It has been indicated that the anisotropy and
its temperature dependences of the magnetoresistance will give
information on the degree of one-dimensionality and the phase
relaxation time.  }

\kword
{
magnetoresistance, weak localization, quasi-one-dimensional system,
organic conductor, Wigner representation
}

\begin{document}
\sloppy
\maketitle

%%% Sec 1 Introduction %%%

\section{Introduction}

Recently, many experiments have reported metallic properties of highly
conducting doped polymers (HCDP), \mbox{\it e.g.} polyacetylene doped with
iodine,\cite{Reghu} {\it p}-phenylenevinylene doped with sulfuric
acid,\cite{Ahlskog} etc. It is expected that HCDP shows
three-dimensional conductivity when polymer chains are entangled at
random, while quasi-one-dimensional conductivity is expected when they
are well aligned each other. Actually, there are some experiments
which have tried to examine the dimensionality of conduction of the
tensile drawn ($\sim \!1000 \%$) samples of HCDP films by the
measurement of magnetoresistance (MR) at low temperature.
\cite{Reghu,Ahlskog}In these experiments the conductivities were
anisotropic which were analyzed based on the formula for anisotropic
three-dimensional systems. However, the resulting anisotropy turned out
to be very large, which invalidate the original assumption of
anisotropic three-dimensionality, \mbox{i.e.} the closed Fermi surface 
with the anisotropic mass. Instead, the results seem to indicate
that the Fermi surface is open for which there have been few
theoretical studies on
MR.\cite{Nakhmedov,Dupuis1,Dupuis2,Dorin,Cassam,Mauz}

In this paper, the weak field MR for such systems with open Fermi
surfaces are theoretically studied by use of the Wigner
representation.

The field theoretical studies of weak-localization (WL)
effects\cite{WLrev} on MR have discussed by Hikami \mbox{\it et
al. }\cite{Hikami} and Kawabata\cite{Kawabata} for two- and
three-dimensional metallic conductors, respectively. In these studies
where the closed Fermi surfaces are assumed the quantum corrections to
the conductivity given by the Cooperon propagators have been easily
calculated even in the presence of a magnetic field in terms of the
Landau quantization. For systems with open Fermi surfaces, on the other 
hand, the eigenvalues of the Cooperon propagator can not be explicitly
given. In order to overcome this difficulty and to study MR
systematically we make use of the Wigner representation.
 
In \S 2 a brief review of the preceding theory for three-dimensional
systems is given, and studies of quasi-one-dimensional systems by the
Wigner representation are given in \S 3. The asymptotic forms of the
MR in three- and one-dimensional limit and summary are given in \S 4
and \S 5, respectively.

We take a unit of $\hbar = 1$.

%\vspace{10cm}

%%% Sec 2 Magnetoresistance in Three dimensional Systems %%%
%%%% SEC 2 MAGNETORESISTAMCE IN THREE DIMENSIONAL SYSTEMS %%%%%%%

\section{Magnetoresistance in Three-Dimensional Systems}

\begin{figure}[hbtp]
%\figureheight{3cm}
\begin{center}
\epsfile{file=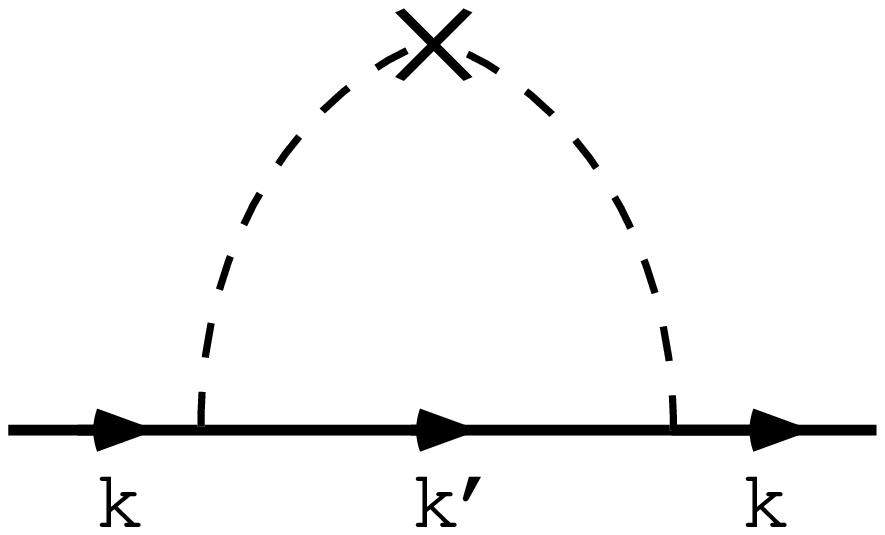,width=5cm}
\caption{Self-energy correction due to the impurity scattering.}
\label{tauz}
\end{center}
\end{figure}

  For three-dimensional systems, we take the model Hamiltonian,
\begin{equation}
{\cal H} = \frac{\bvec{p}^2}{2m} + u \sum_{l} \delta(\bvec{r} -\bvec{R}_l),
\end{equation}
where $u$ is the strength of the short range impurity potential and
$\bvec{R}_l$ is the impurity site. We will consider the quantum
correction term for the conductivity in the order of $(\EF
\tauz)^{-1}$, where $\EF$ is the Fermi energy and $\tauz$ is the
relaxation time due to elastic scattering by impurities given in
\mbox{Fig.} \ref{tauz}. In this figure dashed lines and a cross represent
impurity potentials and the averaging procedure over the distribution of
impurities. This $\tauz$ is given as follows,
\begin{equation}
\tauz^{-1} = 2 \pi n_i u^2 N(0),
\end{equation} 
where $n_i$ is the density of impurities and $N(0)$ is the density of
state per spin at the Fermi energy. 

%%%Figure of relaxation time%%%%%%%%

%%%Figure of WL correction%%%%%%
\begin{figure}[hbtp]
%\figureheight{3cm}
\begin{center}
\epsfile{file=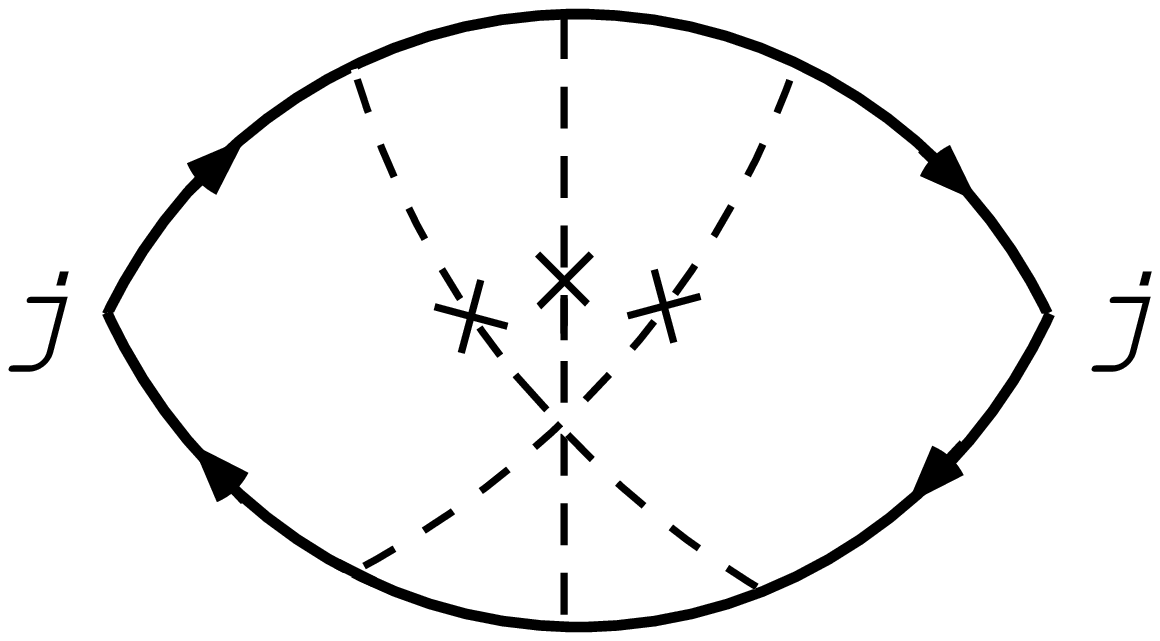,width=5.5cm}
\caption{Weak-localization correction due to the ``Cooperon''.}
\label{WL}
\end{center}
\end{figure}

%%%Figure of the Cooperon%%%%%%
\begin{fullfigure}[hbtp]
%\figureheight{3cm}
\begin{center}
\epsfile{file=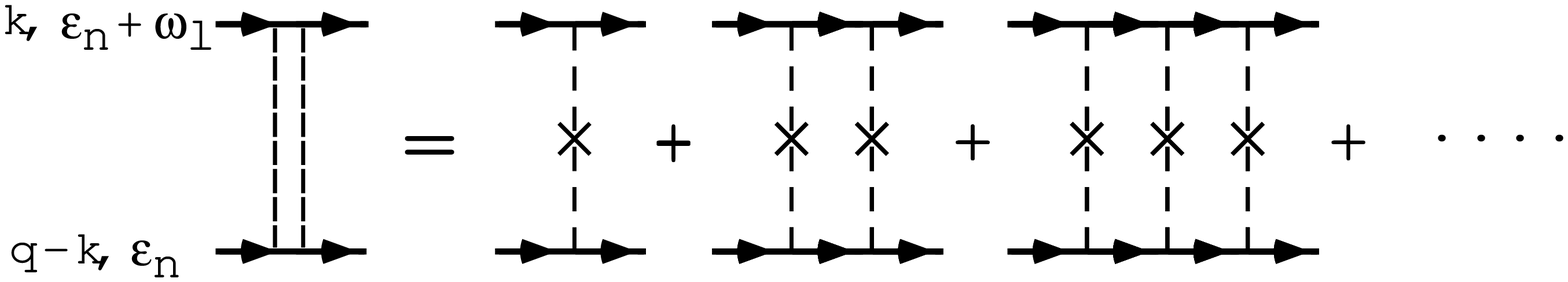,width=12.5cm}
\caption{The Cooperon representing the quantum interference effect.}
\label{CooperonFig}
\end{center}
\end{fullfigure}
 
  The weak-localization effect can be calculated by the summation of
so-called maximally crossed diagrams as given in \mbox{Fig.} \ref{WL}. 
In these diagrams the ladder part (see \mbox{Fig.} \ref{CooperonFig})
which is called the ``Cooperon'' represents the quantum interference
effect between two electrons having nearly opposite wave number. The
Cooperon is singular when $\vare_n (\vare_n + \omega_l) < 0$ where
$\vare_n = (2n+1)\pi k_{\rm B}T$, $\omega_l = 2l\pi k_{\rm B}T$ and
$k_{\rm B}$ is Boltzmann constant, and in this case it is written as
follows,
\begin{equation}
D_c(\vecq,\omega_l) = \frac{1}{2\pi N(0) \tauz^2}
\frac{1}{D\vecq^2 + |\omega_l| +1/\taue},
\end{equation}
where $D=2\EF \tauz/3m$ is the diffusion constant and $\taue$ is the
phase relaxation time due to inelastic scattering introduced
phenomenologically. Then the quantum correction to the conductivity
(\mbox{Fig.} \ref{WL}) is given by
\begin{equation} 			
\frac{\Delta \sigma}{\sigma_0} = -2 \ \tauz^2 \ \Tr \ D_c(\vecq,0) ,
\end{equation}
where $\sigma_0 = 2 e^2 N(0) D $ is the Drude conductivity and Tr
means quantum mechanical trace, \mbox{\it e.g.} $\displaystyle{\sum_q}$ in
the absence of the magnetic field.
  
In the presence of a magnetic field, $H$, whose strength is not so
strong, in the sense $\omega_c \equiv eH/mc \ll \tauz^{-1}$, its effects
can be treated quasiclassically, \mbox{i.e.} $\bvec{q}$ in the
Cooperon is replaced by $\bvec{q}+2e\bvec{A}/c \equiv \bvec{\pi}$,
where $\bvec{A}$ is a vector potential. Fortunately, the Cooperon
depends only on $\bvec{\pi}^2$, so that the trace can easily be
carried out by the use of the eigenstates of Landau
quantization. Hence, the quantum correction is given as
follows,\cite{Kawabata}
\begin{eqnarray}
\label{3dMR}
\frac{\Delta \sigma(H)}{\sigma_0} 
& = & - \frac{1}{2\pi^3 N(0) \ell^2} \nonumber \\ 
& & \times \sum_{N}
 \int {\rm d}q_z \frac{1}{\displaystyle \frac{4D}{\ell^2}\left (
 N+\frac{1}{2}\right) + Dq_z^2 + 1/\taue},
\end{eqnarray}
where $\ell = \sqrt{c/eH}$ is the Larmor radius. Equation
(\ref{3dMR}) is valid for both weak and strong magnetic field limit,
\mbox{i.e.}\  $\ell \gg \Le \equiv \sqrt{D \taue}$\  and \ $ \ell \ll
\Le$, as long as  the conditions, $\omega_c \ll \EF$\  and \ $\ell \gg
\sqrt{D\tauz}$, are satisfied. Especially for weak magnetic field, we
get the following asymptotic form of the magnetoconductance,
$\delta \sigma(H) \equiv \Delta\sigma(H) - \Delta\sigma(0)$,
\begin{equation}
\label{isotropic}
\frac{\delta \sigma(H)}{\sigma_0} = \frac{1}{24\pi^2 N(0)} 
\frac{\sqrt{D} \taue^{3/2}}{\ell^4} \ \propto \ H^2.
\end{equation}

If we assume the anisotropic mass, $m_i \ (i = x, y, z)$, the diffusion
constants are defined as $D_i = 2 \EF \tauz / 3 m_i$, and  
$\delta \sigma(H)$ along the symmetry axis is rewritten as 
\begin{equation}
\label{anisotropic}
\frac{\delta \sigma(H)}{\sigma_0} = \frac{1}{24\pi^2 N(0)} 
\frac{D_1}{\sqrt{D_2}} \frac{\taue^{3/2}}{\ell^4},
\end{equation}
where $D_1$ is the geometric mean of $D_i$s perpendicular to the
magnetic field and $D_2$ is the diffusion constant of the direction of 
magnetic field.

%%% Sec 3 Quasi-One-Dimensional Systems %%%

\section{Magnetoresistance in Quasi-One-Dimensional Systems}

Now, we turn to our problem of quasi-one-dimensional systems with open
Fermi surfaces. We take the model Hamiltonian,
\begin{equation}
{\cal H} = \frac{p_{z}^2}{2m} - \alpha(\cos p_{x} d +
\cos p_{y} d) + u \sum_{l} \delta(\bvec{r} - \bvec{R}_l),
\end{equation}
where $z$-axis is the polymer chain axis, $\alpha$ is the band width
due to the transverse hopping of electrons among chains and $d$ is
the lattice spacing perpendicular to the chain direction. The
one-particle thermal Green function is given as
\fulltext
\begin{equation}
G(\bvec{k}, {\rm i}\vare_n) = 
\frac{1}{{\rm i}\vare_n - \left[ k_z^2/2m - 
\alpha(\cos k_x d + \cos k_y d) - \EF \right] +
{\rm i\, sgn}(\vare_n)/2 \tauz}.
\end{equation}
If the Fermi energy $\EF$ is large enough compared to the band width
in the perpendicular directions, $\alpha$, which is assumed throughout
this paper, and then the warping of the Fermi surface can be ignored
in the integration of a single particle Green function, the relaxation
time due to impurity scattering is given by
\begin{equation}
\tauz^{-1} = \frac{2 n_i u^2}{d^2 \VF} ,
\end{equation}
where $\VF = \sqrt{2\EF/m}$. On the other hand, the cosine band
structure has to be properly treated in the derivation of the Cooperon
as follows,
\begin{equation}
D_c(\vecq,\omega_l) = \frac{n_i u^2}{1 - n_i u^2 X(\vecq, \omega_l)} ,
\end{equation}
\begin{eqnarray}
X(\vecq, \omega_l) & = & \int\! \frac{{\rm d}\bvec{k}}{(2\pi)^3}
\ G(\bvec{k}, {\rm i}\vare_n + {\rm i}\omega_l)
\  G(\vecq-\bvec{k}, {\rm i}\vare_n) \nonumber\\
& = & \int\! \frac{{\rm d}\bvec{k}}{(2\pi)^3} 
\frac{1}{{\rm i}(\vare_n+\omega_l)-
\left\{ k_z^2/2m - \alpha(\cos k_x d + \cos k_y d) - \EF
\right\} + {\rm i}/2 \tauz} \nonumber \\
& & \times \frac{1}{{\rm i}\vare_n -
\left\{ (q_z-k_z)^2/2m - \alpha\left[
\cos(q_x-k_x)d + \cos(q_y-k_y)d\right] - \EF
\right\} - {\rm i}/2 \tauz},
\end{eqnarray}
where $X(\vecq, \omega_l)$ is the polarization function.  The result
of the integration with respect to $k_z$ is given as
follows under the conditions, $\EF \gg \alpha$ and 
$1 \gg \alpha\tauz|\sin\frac{q_{x,y}d}{2}|$, \, $\VF^2\tauz^2
q_z^2$, \, $|\omega_l|\tauz$, 
\begin{eqnarray}
X(\vecq,\omega_l) & \simeq & \frac{2}{\VF} \int\! 
\frac{{\rm d}k_x {\rm d}k_y}{(2\pi)^2} 
%\nonumber \\ & & \times 
\frac{1}{\displaystyle{
\omega_l+\frac{1}{\tauz}
 + \VF^2 \tauz q_z^2 +
2{\rm i}\alpha\left[ \sin\frac{(2k_x-q_x)d}{2} \sin\frac{q_x d}{2} +
\sin\frac{(2k_y - q_y)d}{2} \sin\frac{q_y d}{2}\right] }} \nonumber\\
& \simeq & \frac{2 \tauz}{\VF d^2}\left[ 
1 - 2\alpha^2 \tauz^2 
\left(\sin^2\frac{q_x d}{2}+\sin^2\frac{q_y d}{2} \right) - 
\VF^2\tauz^2 q_z^2 - |\omega_l|\tauz \right] . 
\end{eqnarray}
Then the Cooperon is obtained as\cite{Prigodin1,Prigodin2,Abrikosov}
\begin{equation}
\label{Cooperon}
D_c(\vecq,\omega_l) = \frac{d^2 \VF}{2 \tauz^2} 
\frac{1}{\VF^2 \tauz q_z^2 + \alpha^2 \tauz (2 - \cos q_x d - \cos q_y
d) + |\omega_l| + 1/\taue}.
\end{equation}
\halftext

The quantum corrections to the conductivity (\mbox{Fig.} \ref{WL}) for
each direction under the same conditions as in the
derivation of the Cooperon, \mbox{eq.} (\ref{Cooperon}), are as follows,
\begin{subeqnarray}
\label{QCforQ1D}
\frac{\Delta \Spara}{\Spara} & = & -2 \ \tauz^2 \ \Tr \ D_c(\vecq,0) ,  \\
\frac{\Delta \Sperp}{\Sperp} & = & -2 \ \tauz^2 \ \Tr \cos{q_x d}
\ D_c(\vecq,0).
\end{subeqnarray}
In these equations, the classical conductivities
for each direction are given by
\begin{subeqnarray}
\Spara & = & 2 e^2 N(0) \Dpara , \\
\Sperp & = & 2 e^2 N(0) \Dperp ,
\end{subeqnarray}
where the symbols $\|$ and $\bot$ represent the directions parallel
and perpendicular to the chain axis, respectively, which will be used
in the following as well. Here the density of state and the
diffusion constants are defined as follows,
\begin{subeqnarray}
\label{definition}
N(0) & = & \frac{1}{\pi d^2 \VF} ,\\
\Dpara & = & \VF^2 \tauz , \\
\Dperp & = & \frac{1}{2} \alpha^2 d^2 \tauz ,
\end{subeqnarray}
which are deduced from \mbox{eq.} (\ref{Cooperon}) in the continuum limit,
\mbox{$d \rightarrow 0$}.

%%%Figure of Fermi Surface%%%%%%%%
\begin{figure}[hbtp]
%\figureheight{3cm}
\begin{center}
\epsfile{file=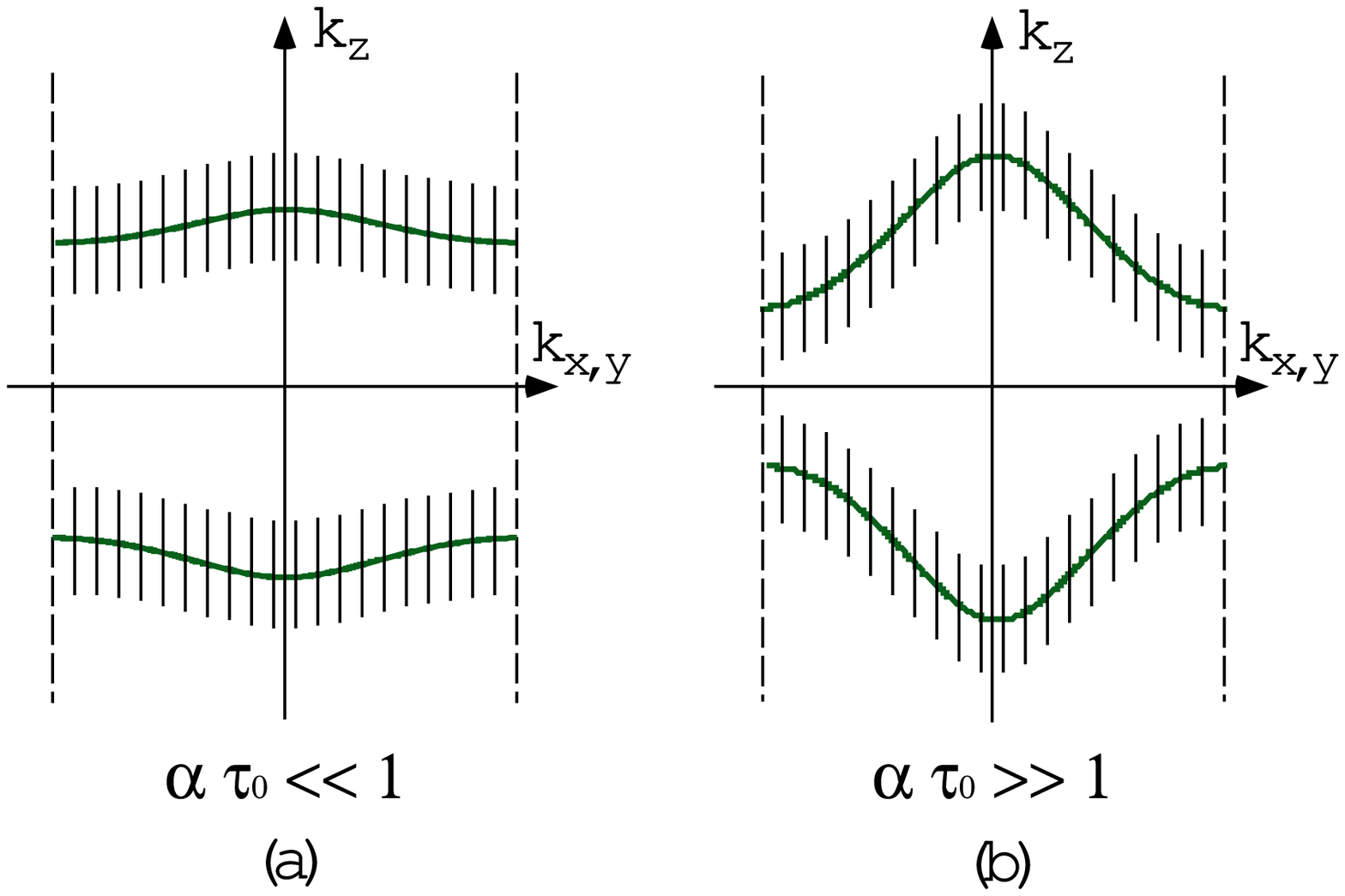,width=7cm}
\caption{Fermi surfaces in the cases, $\alpha \tauz
\ll 1$, (a), and $\alpha \tauz \gg 1$, (b). Here the vertical 
lines represent the broadening of the Fermi surface corresponding to
the energy width, $\tauz^{-1}$.}
\label{Fermi}
\end{center}
\end{figure}

In the limit, $\alpha \tauz \ll 1$, where the warping of the Fermi
surface is less than the broadening, $\tauz^{-1}$, (see \mbox{Fig.} 
\ref{Fermi} (a)), the conditions, $1 \gg \alpha \tauz |\sin
\frac{q_{x,y}d}{2}|$, are satisfied over the whole
Brillouin zone, hence any cutoff is not necessary in the integrations
with respect to $q_x$ and $q_y$ in the evaluations of \mbox{eq.} 
(\ref{QCforQ1D}). On the other hand, in the limit, $\alpha\tauz \gg
1$, where the warping of the Fermi surface is larger than the
broadening, $\tauz^{-1}$,
(see \mbox{Fig.} \ref{Fermi} (b)), the conditions, $1 \gg
\alpha\tauz|\sin\frac{q_{x,y}d}{2}|$, required to derive \mbox{eqs.} 
(\ref{Cooperon}) and (\ref{QCforQ1D}) imply $|q_{x,y}| \ \lsim \
(\alpha \tauz d)^{-1}$.  In this case, however, the main contributions
to the quantum corrections, \mbox{eq.} (\ref{QCforQ1D}), turn out to
be given by the small $q$ such as $|q| \ \lsim \ (\alpha
\sqrt{\tauz\taue} d)^{-1}$ due to the lifetime of the Cooperon,
$\taue$. Since $(\alpha \sqrt{\tauz\taue} d)^{-1} < (\alpha
\tauz d)^{-1}$ is usually satisfied (\mbox{i.e.} $\taue \gg \tauz$),
the present estimations of the quantum corrections based on  
\mbox{eqs.} (\ref{Cooperon}) and (\ref{QCforQ1D}) are justified even
in this case of $\alpha\tauz \gg 1$.

To obtain the MR, we replace $\bvec{q}$ by
$\bvec{\pi}=\bvec{q}+2e\bvec{A}/c$ ,
\begin{subeqnarray}
\label{trace}
\frac{\Delta \Spara}{\Spara} = - {\rm Tr} \frac{d^2 \VF}{\Dpara
\pi_z^2 + \alpha^2 \tauz (2 - \cos \pi_x d - \cos \pi_y d) + 1/\taue
}, \\
\frac{\Delta \Sperp}{\Sperp} = - {\rm Tr} \frac{d^2 \VF \cos \pi_x d}
{\Dpara \pi_z^2 + \alpha^2 \tauz (2 - \cos \pi_x d - \cos \pi_y d)
+ 1/\taue}.
\end{subeqnarray}
Here we must be careful to treat $\pi_i$s because of their
noncommutability,
\begin{equation}
[ \pi_i , \pi_j ] = {\rm i} \frac{2}{\ell^2}\vare_{ijk},
\end{equation}
where $k$ is the direction of magnetic field and $\vare_{ijk}$ is
Levi-Civita's totally antisymmetric tensor. Since $\pi_i$s are
contained in cosine terms in our Cooperon, we cannot use the Landau
quantization method and it is impossible to study MR for arbitrary
field. However for studies in weak magnetic field, the method of
Wigner representation\cite{Kubo} is most suited, because it is
a systematic method of expanding physical quantities in terms of the
small parameter which is the value of the commutator of canonical
variables. Moreover as it turned out, the MR in a weak field yields
important information on the degree of the alignment of the polymer
and the phase relaxation time.

In the Wigner representation, the trace of some physical quantity,
$A(\hat{p},\hat{q})$, which is given as a function of
canonical variables, $\hat{p}$ and $\hat{q}$ satisfying $[\,\hat{p}\: ,
\:\hat{q}\,]=-{\rm i}c$, can be obtained by replacing quantum 
operators to corresponding classical differential operators operating
on $1$, and integrating the quantity over classical variables, $p$ and
$q$,
\begin{equation}
\label{Wigner}
\Tr \ A(\hat{p},\hat{q}) = \frac{1}{2\pi c}\int\! {\rm d}p{\rm d}q 
A\left(p+\frac{c}{2{\rm i}}\del{}{q} \:,\:
q-\frac{c}{2{\rm i}}\del{}{p}\right) \cdot 1 .
\end{equation}

%%%Figure of possible configurations%%%%%%%%
\begin{figure}[hbtp]
%\figureheight{3cm}
\begin{center}
\epsfile{file=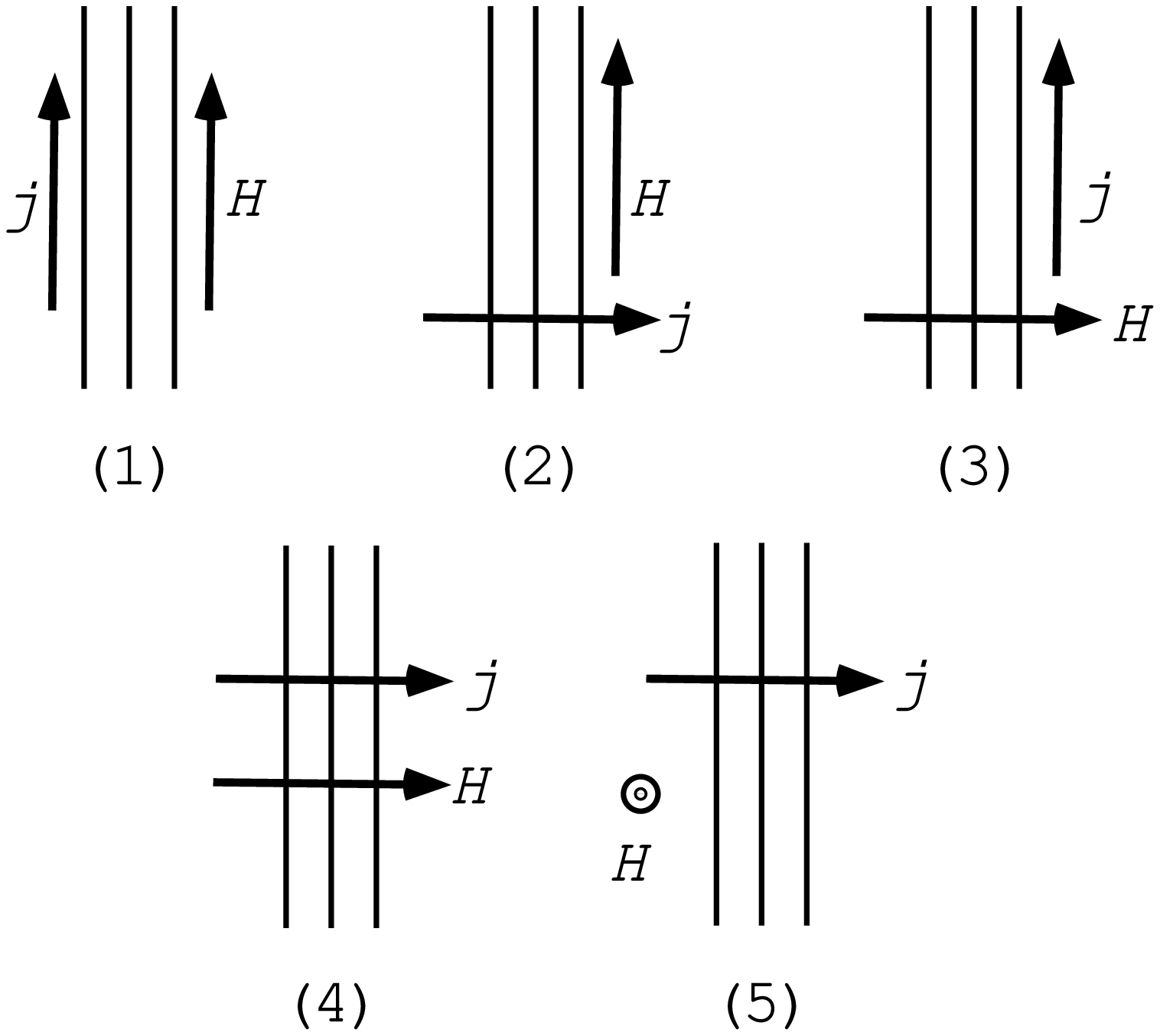,width=7cm}
\caption{Five possible configurations of current and field with
respect to the chain direction in the measurement of MR.}
\label{configurations}
\end{center}
\end{figure}

In our case of MR in quasi-one-dimensional systems, two components of
$\hat{\pi}$ perpendicular to the magnetic field correspond to
$\hat{p}$ and $\hat{q}$ in \mbox{eq.} (\ref{Wigner}). For each of the
five possible configurations as shown in
\mbox{Fig.} \ref{configurations} we have to replace the operators as 
follows,
\begin{equation}
\hat{\bvec{\pi}} \rightarrow \bvec{\pi}+
\frac{1}{{\rm i}\ell^2} \bvec{h} \times \del{}{\bvec{\pi}},
\end{equation}
where $\bvec{h}$ is the unit vector along the direction of magnetic
field, and integrate over $\bvec{\pi}$. For example the quantum
correction in the
\mbox{config.} (1) in Fig.~\ref{configurations}, we have to evaluate 
the following,
\fulltext
\begin{equation}
\frac{\Delta \sigma_1}{\Spara} = - \frac{d^2 \VF \taue}{(2\pi)^3}
\int_0^{\infty}\!\!\!\!{\rm d}s \int\!{\rm d}^3\pi\ \mbox{\Large e}^{
-s \left\{ \Dpara \taue \pi_z^2 + \alpha^2 \tauz \taue \left[2 -
\cos \big(\pi_x-\frac{1}{{\rm i} \ell^2}\del{}{\pi_y}\big) d - 
\cos \big(\pi_y+\frac{1}{{\rm i} \ell^2}\del{}{\pi_x}\big) d 
\right] + 1 \right\}} \cdot 1,
\end{equation}
%\end{full}
where the integrations with respect to $\pi_x$ and $\pi_y$ can be taken
over the whole Brillouin zone.

The explicit evaluations of the quantum corrections up to the second
order of $H$ for each configuration result in as follows,
%\begin{full}
\begin{equation}
\label{q1d}
\begin{array}{rlcclr}
{\displaystyle
\frac{\Delta \sigma_1}{\Spara} } \ = 
& \hspace{-2mm} {\displaystyle -
\frac{1}{2\sqrt{\pi}}\sqrt{\frac{\taue}{\tauz}}
\int_{0}^{\infty} {\rm d}s\  
\mbox{\Large e}^{-s(4a+1)}
\ [ \ s^{\mbox{\tiny $-1/2$}}\ \I_{0}(2as)^2\ }  
& \hspace{-2mm} 
{\displaystyle 
 - \ \frac{2}{3} } 
\hspace{-2mm} & \hspace{-2mm}
{\displaystyle 
\left( \frac{\Leperp^2}{\ell^2} \right)^2 } 
\hspace{-2mm} & \hspace{-2mm}
{\displaystyle s^{\mbox{\tiny $3/2$}}\ \I_{1}(2as)^2\ } \hspace{-2mm}
& \hspace{-2mm} ], \\ \\
{\displaystyle \frac{\Delta \sigma_2}{\Sperp}} \ = 
& \hspace{-2mm}
{\displaystyle - 
\frac{1}{2\sqrt{\pi}}\sqrt{\frac{\taue}{\tauz}}
\int_{0}^{\infty} {\rm d}s\ 
\mbox{\Large e}^{-s(4a+1)}
\ [\ s^{\mbox{\tiny $-1/2$}}\ \I_{0}(2as)\ \I_{1}(2as)\ }  
& \hspace{-2mm} 
{\displaystyle 
- \ \frac{2}{3} } 
\hspace{-2mm} & \hspace{-2mm}
{\displaystyle
\left( \frac{\Leperp^2}{\ell^2} \right)^2 } 
\hspace{-2mm} & \hspace{-2mm}
{\displaystyle 
s^{\mbox{\tiny $3/2$}}\ \I_{0}(2as)\ \I_{1}(2as)\ } \hspace{-2mm} & \hspace{-2mm} ], \\ \\
{\displaystyle \frac{\Delta \sigma_3}{\Spara}} \ = 
& \hspace{-2mm}
{\displaystyle - 
\frac{1}{2\sqrt{\pi}}\sqrt{\frac{\taue}{\tauz}}
\int_{0}^{\infty} {\rm d}s\  
\mbox{\Large e}^{-s(4a+1)}
\ [ \ s^{\mbox{\tiny $-1/2$}}\ \I_{0}(2as)^2\ }  
& \hspace{-2mm} 
{\displaystyle 
- \ \frac{2}{3}} 
\hspace{-2mm} & \hspace{-2mm}
{\displaystyle  
\left( \frac{\Leperp\Lepara}{\ell^2} \right)^2 } 
\hspace{-2mm} & \hspace{-2mm} 
{\displaystyle 
s^{\mbox{\tiny $3/2$}}\ \I_{0}(2as)\ \I_{1}(2as)\ } \hspace{-2mm} & \hspace{-2mm} ],  \\ \\
{\displaystyle \frac{\Delta \sigma_4}{\Sperp}} \ = 
& \hspace{-2mm}
{\displaystyle - 
\frac{1}{2\sqrt{\pi}}\sqrt{\frac{\taue}{\tauz}}
\int_{0}^{\infty} {\rm d}s\  
\mbox{\Large e}^{-s(4a+1)}
\ [ \ s^{\mbox{\tiny $-1/2$}}\ \I_{0}(2as)\  \I_{1}(2as)}  
& \hspace{-2mm} 
{\displaystyle 
- \ \frac{2}{3} } 
\hspace{-2mm} & \hspace{-2mm} 
{\displaystyle 
\left( \frac{\Leperp\Lepara}{\ell^2} \right)^2 } 
\hspace{-2mm} & \hspace{-2mm}
{\displaystyle  
s^{\mbox{\tiny $3/2$}}\ \I_{1}(2as)^2\ } \hspace{-2mm} & \hspace{-2mm} ],  \\ \\
{\displaystyle \frac{\Delta \sigma_5}{\Sperp}} \ = 
& \hspace{-2mm}
{\displaystyle - 
\frac{1}{2\sqrt{\pi}}\sqrt{\frac{\taue}{\tauz}}
\int_{0}^{\infty} {\rm d}s\  
\mbox{\Large e}^{-s(4a+1)}
\ [ \ s^{\mbox{\tiny $-1/2$}}\ \I_{0}(2as)\  \I_{1}(2as)}  
& \hspace{-2mm} 
{\displaystyle 
- \ \frac{2}{3} } 
\hspace{-2mm} & \hspace{-2mm} 
{\displaystyle 
\left( \frac{\Leperp\Lepara}{\ell^2} \right)^2 } 
\hspace{-2mm} & \hspace{-2mm}
{\displaystyle  
s^{\mbox{\tiny $3/2$}}\ \I_{0}(2as)^2} \hspace{-2mm} & \hspace{-2mm} ],  \\ 
\end{array}
\end{equation}
\halftext
where $\Delta \sigma_i$ is the quantum correction for the $i$-th
configuration in \mbox{Fig.} \ref{configurations}, $\I_{0}(z)$ and 
$\I_{1}(z)$ are the modified Bessel functions, $\Lepara \equiv
\sqrt{\Dpara \taue}$ and $\Leperp \equiv \sqrt{\Dperp \taue}$ are the
phase relaxation lengths for each direction, and
\begin{equation}
a \equiv \frac{1}{2} \alpha^2 \tauz \taue = \left( \frac{\Leperp}{d}
\right)^2
\end{equation}
is the ``dimensionality parameter'' whose meaning is discussed
below. In each of \mbox{eqs.} (\ref{q1d}), the first term in the
integral is the WL correction in the absence of the magnetic field,
$\Delta\sigma_i(0)$, and the second term is the magnetoconductance,
$\delta\sigma_i(H) \equiv \Delta\sigma_i(H) - \Delta\sigma_i(0)$. The
expansion parameters are $\Leperp^2 / \ell^2$ for $H \| z$ and
$\Leperp \Lepara / \ell^2$ for $H \bot z$, respectively. This is
easily understood because the magnetic field always affect electrons
through the orbital motion within the plane perpendicular to the
field.

The parameter, $a$, represents the dimensionality in the sense of the
quantum interference effects due to the Cooperon, and its physical
meaning is how many chains electrons can hop through with their
coherency kept. The interference of electrons is three-dimensional if
$a$ is large, $a \gg 1$, even though the Fermi surface is open because
electrons can move among many chains by diffusive motion until they
lose their phase memory. On the other hand, it is one-dimensional if
$a$ is small, $a
\ll 1$, since electrons cannot keep coherency even in a single
hopping.

%%% Sec 4 The Asymptotic Forms %%%
\section{The Asymptotic Forms}

\fulltext

In this section, the asymptotic forms of the conductivity in three
and one dimensions are elucidated:

\subsection{Three-dimensional limit}

The three-dimensional limit, $a \gg 1$, of \mbox{eqs.} (\ref{q1d}) can
be obtained by using the asymptotic form of modified Bessel function,
and the results are as follows,
\begin{equation}
\label{3d}
\begin{array}{rlcclr}
{\displaystyle \frac{\Delta \sigma_1}{\Spara} \ = } 
& \hspace{-2mm}
{\displaystyle \ - \  
\frac{1}{2\pi \alpha \tauz} \ \left[ \  1.61 \sqrt{\pi} \ - \  
\frac{1}{\alpha \tauz} \sqrt{ \frac{\tauz}{\taue}} \ \right] \  }  
& \hspace{-2mm} 
{\displaystyle 
+ \ \frac{1}{24 \pi} \sqrt{ \frac{\taue}{\tauz}} \  
\left( \frac{d \Leperp}{\ell^2} \right)^2  }, \\ \\

{\displaystyle \frac{\Delta \sigma_2}{\Sperp} \ = }
& \hspace{-2mm}
{\displaystyle \ - \  
\frac{1}{2\pi \alpha \tauz} \ \left[ \  0.41 \sqrt{\pi} \ - \  
\frac{1}{\alpha \tauz} \sqrt{ \frac{\tauz}{\taue}} \ \right] \  } 
& \hspace{-2mm} 
{\displaystyle 
+ \ \frac{1}{24 \pi} \sqrt{ \frac{\taue}{\tauz}} \  
\left( \frac{d \Leperp}{\ell^2} \right)^2  }, \\ \\

{\displaystyle \frac{\Delta \sigma_3}{\Spara} \ = } 
& \hspace{-2mm}
{\displaystyle \ - \  
\frac{1}{2\pi \alpha \tauz} \ \left[ \  1.61 \sqrt{\pi} \ - \  
\frac{1}{\alpha \tauz} \sqrt{ \frac{\tauz}{\taue}} \ \right] \  } 
& \hspace{-2mm} 
{\displaystyle 
+ \ \frac{1}{24 \pi} \sqrt{ \frac{\taue}{\tauz}} \  
\left( \frac{d L_{\vare \mbox{\tiny $\|$}}}{\ell^2} \right)^2  }, \\ \\

{\displaystyle \frac{\Delta \sigma_4}{\Sperp} \ = } 
& \hspace{-2mm}
{\displaystyle \ - \  
\frac{1}{2\pi \alpha \tauz} \ \left[ \  0.41 \sqrt{\pi} \ - \  
\frac{1}{\alpha \tauz} \sqrt{ \frac{\tauz}{\taue}} \ \right] \  } 
& \hspace{-2mm} 
{\displaystyle 
+ \ \frac{1}{24 \pi} \sqrt{ \frac{\taue}{\tauz}} \  
\left( \frac{d \Lepara}{\ell^2} \right)^2  }, \\ \\

{\displaystyle \frac{\Delta \sigma_5}{\Sperp} \ = } 
& \hspace{-2mm}
{\displaystyle \ - \  
\frac{1}{2\pi \alpha \tauz} \ \left[ \  0.41 \sqrt{\pi} \ - \  
\frac{1}{\alpha \tauz} \sqrt{ \frac{\tauz}{\taue}} \ \right] \  } 
& \hspace{-2mm} 
{\displaystyle 
+ \ \frac{1}{24 \pi} \sqrt{ \frac{\taue}{\tauz}} \  
\left( \frac{d \Lepara}{\ell^2} \right)^2  }. \\ 
\end{array}
\end{equation} 
These are identical with the conclusions of preceding
theories of WL and weak field MR in three-dimensional systems,
\cite{WLrev,Hikami,Kawabata,Prigodin1,Prigodin2,Abrikosov}
with the density of state, $N(0)$, and the anisotropic tensor
components of the diffusion constants, $\Dpara$ and $\Dperp$, as given
in \mbox{eq.} (\ref{definition}), \mbox{\it e.g.} the substitution of
them for \mbox{eq.} (\ref{anisotropic}) gives the second terms,
$\delta \sigma_i(H)$, of \mbox{eqs.} (\ref{3d}).  This is expected
because in the limit, $\alpha^2 \tauz \taue \gg 1$, the main
contribution to the integration of the Cooperon is given by small $q$
such as $|q_x|, |q_y| \ \lsim \ (\alpha \sqrt{\tauz\taue} d)^{-1}$.
Therefore our formulae, \mbox{\it e.g.} \mbox{eqs.} (\ref{Cooperon}) and
(\ref{QCforQ1D}), turn out to be the same as those in anisotropic
three-dimensional systems shown in \S 2. This is the reason why the
quantum corrections of the systems with $\alpha\tauz \gg 1$ are given
by those of the anisotropic three-dimensional systems even though the
Fermi surface is open, since $\alpha^2 \tauz \taue \gg 1$ because of 
$\taue \gg \tauz$.

%\halftext

\subsection{One-dimensional limit}

When the system becomes one-dimensional, $a \ll 1$, the asymptotic
forms are given as
\begin{equation}
\label{1d}
\begin{array}{rlllll}
{\displaystyle \frac{\Delta \sigma_1}{\sigma_{\mbox{\tiny $\|$}}} \ = } 
& \hspace{-1mm}
{\displaystyle \ - \  
\frac{1}{2} \ \sqrt{ \frac{\tau_{\vare}}{\tau_0}} \  }  
& \hspace{-1mm} 
{\displaystyle 
+ \  \ \frac{35}{16} \ \sqrt{ \frac{\tau_{\vare}}{\tau_0}} }
\hspace{-1mm} & {\displaystyle a^4} & \hspace{-3mm}
{\displaystyle    
\left( \frac{d^2}{\ell^2} \right)^2  }, \\ \\

{\displaystyle \frac{\Delta \sigma_2}{\sigma_{\mbox{\tiny $\bot$}}} \ = }
& \hspace{-1mm}
{\displaystyle \ - \  
\frac{1}{8} \ \sqrt{ \frac{\tau_{\vare}}{\tau_0}} \ a \ }  
& \hspace{-1mm} 
{\displaystyle 
+ \  \ \frac{5}{8} \ \sqrt{ \frac{\tau_{\vare}}{\tau_0}} }
\hspace{-1mm} & {\displaystyle a^3} & \hspace{-3mm}
{\displaystyle    
\left( \frac{d^2}{\ell^2} \right)^2  }, \\ \\

{\displaystyle \frac{\Delta \sigma_3}{\sigma_{\mbox{\tiny $\|$}}} \ = } 
& \hspace{-1mm}
{\displaystyle \ - \  
\frac{1}{2} \ \sqrt{ \frac{\tau_{\vare}}{\tau_0}} \  }  
& \hspace{-1mm} 
{\displaystyle 
+ \  \ \frac{5}{8} \ \sqrt{ \frac{\tau_{\vare}}{\tau_0}} }
\hspace{-1mm} & {\displaystyle a^2} & \hspace{-3mm}
{\displaystyle    
\left( \frac{d L_{\vare \mbox{\tiny $\|$}}}{\ell^2} \right)^2  }, \\ \\

{\displaystyle \frac{\Delta \sigma_4}{\sigma_{\mbox{\tiny $\bot$}}} \ = } 
& \hspace{-1mm}
{\displaystyle \ - \  
\frac{1}{8} \ \sqrt{ \frac{\tau_{\vare}}{\tau_0}} \ a \ }  
& \hspace{-1mm} 
{\displaystyle 
+ \  \ \frac{35}{16} \ \sqrt{ \frac{\tau_{\vare}}{\tau_0}} }
\hspace{-1mm} & {\displaystyle  a^3} & \hspace{-3mm}
{\displaystyle    
\left( \frac{d L_{\vare \mbox{\tiny $\|$}}}{\ell^2} \right)^2  }, \\ \\

{\displaystyle \frac{\Delta \sigma_5}{\sigma_{\mbox{\tiny $\bot$}}} \ = } 
& \hspace{-1mm}
{\displaystyle \ - \  
\frac{1}{8} \ \sqrt{ \frac{\tau_{\vare}}{\tau_0}} \ a \ }  
& \hspace{-1mm} 
{\displaystyle 
+ \ \ \ \frac{1}{4} \ \sqrt{ \frac{\tau_{\vare}}{\tau_0}} }
\hspace{-1mm} & {\displaystyle a} & \hspace{-3mm}
{\displaystyle    
\left( \frac{d L_{\vare \mbox{\tiny $\|$}}}{\ell^2} \right)^2  }. \\
\end{array}
\end{equation} 
\halftext
As is easily seen, the second term of \mbox{config.} (1), $\delta
\sigma_1(H)$, will be reduced most rapidly as $a \rightarrow 0$, while
 that of \mbox{config.} (5), $\delta \sigma_5(H)$, will remain larger
than the others.

Hence, one can infer the value of the dimensionality parameter, $a$,
experimentally by the comparison of the anisotropy of the
magnetoconductance, $\delta
\sigma(H)$. For example, the ratio of $\delta \sigma_3(H)$ and $\delta 
\sigma_4(H)$ will give the value of $a$, yielding important
information about the degree of the alignment of polymer.

In addition, the temperature dependence of $a$ thus deduced gives
information on that of the phase relaxation time, $\taue$.

%%% Sec 5 Summary %%%

\section{Summary}

We have developed a theory of weak field MR in quasi-one-dimensional
systems which have open Fermi surfaces. Even though the effects of
magnetic field on electrons with such open Fermi surface are not easy
to treat, the correct results in weak field regime have been
determined by use of the Wigner representation. It is to be noted
that this is a rare case in which the Wigner representation is applied
to a explicit calculation of the quantum transport phenomena.

We have obtained the asymptotic forms of conductivities in three- and
one-dimensional limit in the sense of the quantum interference effect.
We have pointed out that the dimensionality parameter, $a$, and thus
the degree of the alignment of polymers can be inferred by studying
the anisotropy of the magnetoconductance, $\delta \sigma(H)$, for five
possible configurations.  Moreover, the temperature dependence of the
phase relaxation time can be deduced from that of the dimensionality
parameter.

In a more detailed comparison with the experiments, however, the
existence of the mutual interaction effects has to be taken into
account.\cite{e-e} The Coulomb interaction associated with the spin
Zeeman effect gives contributions to MR of the same order as the WL,
 but its sign is opposite and the scaling fields are different; 
\mbox{i.e.} $g\mu_{\rm B}H/k_{\rm B}T$ where $g$ is the Land\'e 
$g$-factor and $\mu_{\rm B}$ is the Bohr magneton in the case of the
interaction effects while $\Leperp^2 / \ell^2$ for $H \| z$ and
$\Leperp\Lepara / \ell^2$ for $H \bot z$, respectively, in the present
WL effects. Since $\Leperp < \Lepara$ will be naturally satisfied, the
scaling field of the WL effects for $H \| z$ should be larger than
that for $H \bot z$, but the magnitude of these scaling fields
(especially that in the case of $H \| z$) relative to that of
interaction effects is not unique. In the case of \mbox{refs.} 
\citen{Reghu} and \citen{Ahlskog}, the scaling field of the
interaction effects comes between those two of the WL effects and the
interaction effects are almost negligible for $H \bot z$, so that one
can infer the dimensionality parameter, $a$, adequately from $\delta
\sigma(H)$ of $H \bot z$.

\section*{Acknowledgments}
We would like to thank Dr.~Reghu Menon for drawing our interest to
refs.~1 and 2. \mbox{Y.~N.} thanks Hiroshi Kohno and Masakazu Murakami
for valuable discussions. We are indebted to Dr.~Achim Rosch who
kindly informed us of their related work.

\vspace{0.5cm} 
\mbox{\it Note added---}Recently, the MR in the same kind of
quasi-one-dimensional systems has also been studied by C.~Mauz,
A.~Rosch and P.~W\"olfle (\PR \ B {\bf 56} (1997) 10953). They focused
on the cases of $H \bot$chain where the Cooperon is described by
Mathieu's equation and discussed various limiting cases. However,
their results of one-dimensional limit are different from ours because
their approximation is not justified when the coupling between chains
is very weak.


\begin{thebibliography}{99}
\bibitem{Reghu} Reghu~M., K.~V\"{a}kiparta, Y.~Cao and D.~Moses: \PR
\ B {\bf 49} (1994) 16162.
\bibitem{Ahlskog} M.~Ahlskog, Reghu~M., A.~J.~Heeger, T.~Noguchi and
T.~Ohnishi: \PR \ B {\bf 53} (1996) 15529. 
\bibitem{Nakhmedov} \'E.~P.~Nakhmedov, V.~N.~Prigodin and Yu.~A.~Firsov: 
{\rm JETP Lett.} {\bf 43} (1986) 743. 
\bibitem{Dupuis1} N.~Dupuis and G.~Montambaux: \PRL \ {\bf 68} (1992) 357. 
\bibitem{Dupuis2} N.~Dupuis and G.~Montambaux: \PR \ B {\bf 46} (1992) 9603. 
\bibitem{Dorin} V.~V.~Dorin: {\rm Phys. Lett.} A {\bf 183} (1993) 233. 
\bibitem{Cassam} A.~Cassam-Chenai and D.~Mailly: \PR \ B {\bf 52} (1995) 1984. 
\bibitem{Mauz} C.~Mauz, A.~Rosch and P.~W\"olfle: 
\PR \ B {\bf 56} (1997) 10953.
\bibitem{WLrev} For a review, P.~A.~Lee and T.~V.~Ramakrishnan:
Rev. Mod. Phys. \ {\bf 57} (1985) 287.
\bibitem{Hikami} S.~Hikami, A.~I.~Larkin and Y.~Nagaoka: \PTP \ {\bf 63} 
(1979) 707.
\bibitem{Kawabata} A.~Kawabata: \SSC \ {\bf 34} (1980) 431.   
\bibitem{Prigodin1} V.~N.~Prigodin and Yu.~A.~Firsov: {\rm JETP Lett.} {\bf 38} (1984) 284
\bibitem{Prigodin2} V.~N.~Prigodin and S.~Roth: \SM \ {\bf 53} (1993) 237. 
\bibitem{Abrikosov} A.~A.~Abrikosov: \PR \ B {\bf 50} (1994) 1415.
\bibitem{Kubo} R.~Kubo: \JPSJ \ {\bf 19} (1964) 2127.
\bibitem{e-e} {\it Electron-Electron Interactions in Disordered Systems}, ed. A.~L.~Efros and  M.~Pollak (North-Holland, Amsterdam, 1985).
\end{thebibliography}
\end{document}